\documentclass[aps,preprint]{revtex4}
\usepackage{graphicx}
\usepackage{amssymb}

\makeatletter

\begin{document}
\global\long\def\met{\not{\!{\rm E}}_{T}}

\begin{flushright}
ANL-HEP-PR-11-45, NSF-KITP-11-125, NUHEP-TH/11-18
\end{flushright}

\title{Calculation of $Wb\bar{b}$ Production via \\ Double Parton Scattering at the LHC}
\author{ Edmond L. Berger$^{a}$}
\email{berger@anl.gov}
\author{C.~B.~Jackson$^{b}$}
\email{cbjackson@uta.edu}
\author{Seth Quackenbush$^a$}
\email{squackenbush@hep.anl.gov}
\author{Gabe Shaughnessy$^{a,c}$}
\email{g-shaughnessy@northwestern.edu} 
\affiliation{
\mbox{$^a$High Energy Physics Division, Argonne National Laboratory, Argonne, Illinois 60439, USA}  \\ 
\mbox{$^b$Physics Department, University of Texas at Arlington, Arlington, Texas 76019, USA}  \\
\mbox{$^c$Department of Physics and Astronomy, Northwestern University, Evanston, Illinois 60208, USA}
}

\begin{abstract}
We investigate the potential to observe double parton scattering at the Large Hadron Collider  in 
$p p \rightarrow Wb\bar{b} X \to \ell \nu b \bar{b} X$ at 7 TeV.  Our analysis tests the efficacy of several kinematic variables in isolating the double parton process of interest from the single parton process and relevant backgrounds for the first 10 fb$^{-1}$ of integrated luminosity.  These variables are constructed to expose the independent nature of the two subprocesses in double parton scattering, $pp \to \ell \nu X$ and $pp \to b \bar{b} X$.  
We use next-to-leading order perturbative predictions for the double parton and single parton scattering components of $W b \bar{b}$ and for the pertinent  backgrounds.   The next-to-leading order contributions are important for a proper description of some of the observables we compute.  We find that the double parton process can be identified and measured with significance $S/\sqrt B \sim 10$, 
provided the double parton scattering effective cross section $\sigma_{\rm eff} \sim 12$~mb.  
 
\end{abstract}

\maketitle

\section{introduction}
\label{sec:intro}

The successful operation of the Large Hadron Collider (LHC) and its detectors opens a new era in particle physics.  The higher energies and larger luminosities at the LHC make it possible to explore new physics scenarios and to investigate unexplored aspects of established theories such as quantum chromodynamics (QCD).

\begin{figure}[t]
\includegraphics[scale=0.7]{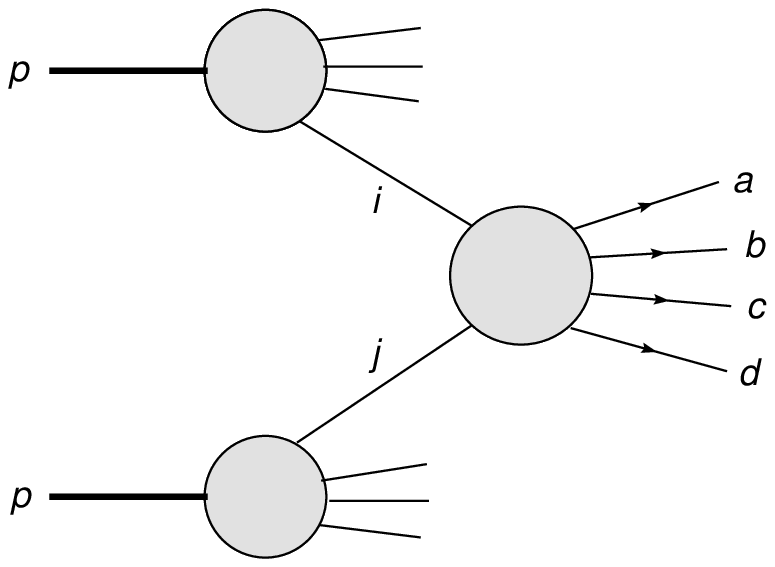}
\hspace{2cm}
\includegraphics[scale=0.7]{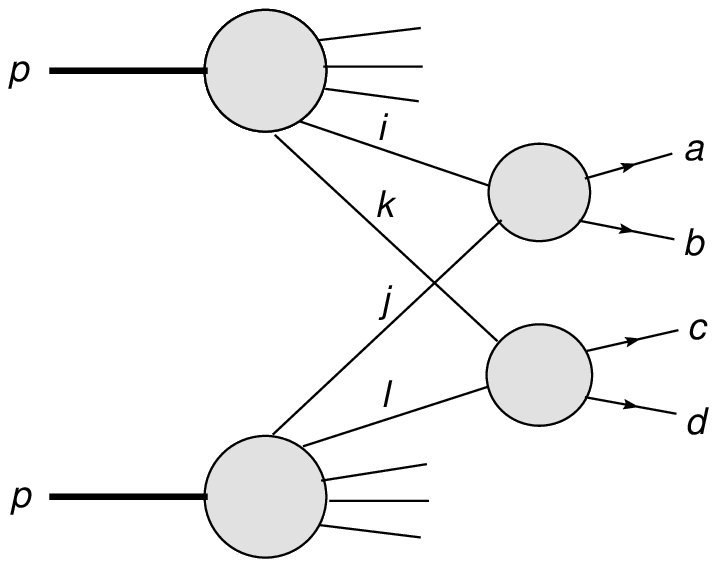}
\caption{Schematic depiction of single parton scattering (left) and double parton scattering (right).  In single parton scattering, one parton from each hadron is active in the scattering and the partonic process is ${\cal{A}}(i j \to a b c d)$.  Double parton scattering assumes two partons from each hadron are active in the hard scattering and the total partonic process consists of two independent subprocesses ${\cal{A}}(i j \to a b)$ and ${\cal{A}}(k \ell \to c d)$. }
\label{fg:sps-dps-cartoons}
\end{figure}

The standard picture of hadron-hadron collisions is shown on the left side of Fig.~\ref{fg:sps-dps-cartoons}.  One parton from each proton partakes in the hard scattering to produce the final state.  The probability density for finding parton $i$ in a proton with momentum fraction $x_i$ and at the factorization scale $\mu$ is parametrized by the parton distribution function (PDF) $f^i_p(x_i, \mu)$.  In this {\it single parton scattering} (SPS) scenario, the differential hadronic cross section neatly factors into: 
\begin{equation}
d\sigma^{SPS}_{pp} = \sum_{i,j} \int f^i_p(x_1,\mu) f^j_p(x_1^\prime, \mu) d\hat{\sigma}_{ij}(x_1,x_1^\prime,\mu) dx_1 dx_1^\prime \,.  
\label{eq:sps-xs}
\end{equation}
The ``short-distance" partonic cross section $d\hat{\sigma}_{ij}$ is computed in perturbation theory, whereas the PDFs are nonperturbative objects and must be extracted from experiment.

This simple picture of proton-proton collisions is incomplete.  The full description of hadronic collisions involves other elements including initial- and final-state soft radiation,  underlying events, and multiparton interactions.  Double parton scattering (DPS) describes the case in which two short-distance subprocesses occur in a given hadronic interaction, with two initial partons being active from each of the incident protons.  The general picture of DPS is shown on the right side of Fig.~\ref{fg:sps-dps-cartoons}.  Given the small probability for single parton scattering in hadronic collisions, it is often assumed that the effects of double (or multiple) parton scattering may be ignored or subsumed into the parametrization of the underlying event.  Nevertheless, it is worth exploring theoretically and investigating experimentally whether a second distinct hard component may be identified in events at the LHC.   Some evidence for DPS has been observed at the CERN Intersecting Storage Rings \cite{Akesson:1986iv}, the CERN Super Proton Synchrotron~\cite{Alitti:1991rd},  and more recently, at the Fermilab Tevatron \cite{Abe:1997xk,D0:2009}.    

In an earlier study~\cite{Berger:2009cm}, we investigated the DPS and SPS contributions at the LHC to the four-parton final state $p p \rightarrow b \bar{b} j j X$ in which  a $b \bar{b}$ system is produced along with two jets $j$.   We showed that there are characteristic regions of phase space in which the DPS events are expected to concentrate, and we developed a methodology to measure the effective size of DPS.  Precise measurements of DPS at the LHC will provide insight into parton correlations, nonperturbative dynamics in hadron-hadron collisions, the structure of the proton, and parton distribution functions.

Theoretically, the study of DPS phenomenology has a long history \cite{Goebel:1979mi,Paver:1982yp,Humpert:1983pw,Mekhfi:1983az,Humpert:1984ay,Ametller:1985tp,Halzen:1986ue,Mangano:1988sq,Godbole:1989ti,Drees:1996rw,Eboli:1997sv,Yuan:1997tr,Calucci:1997uw,Calucci:2009ea,Calucci:2009sv,DelFabbro:1999tf,DelFabbro:2002pw,Kulesza:1999zh,Korotkikh:2004bz,Cattaruzza:2005nu,Hussein:2006xr,Hussein:2007gj,Maina:2009vx,Maina:2009sj,Berger:2009cm,Gaunt:2010pi,Strikman:2010bg,Bandurin:2010gn,Diehl:2011tt,Ryskin:2011kk,Baranov:2011ch,Bartels:2011qi,Calucci:2010wg}.  Under the assumption of weak dynamic and kinematic correlations between the two hard-scattering subprocesses, the general approach in these studies is to assume the differential hadronic cross section takes a factored form in analogy to Eq.~(\ref{eq:sps-xs}):
\begin{eqnarray}
d\sigma^{DPS}_{pp} &=& \frac{m}{2 \sigma_{\rm eff}} \sum_{i,j,k,l} \int H_p^{ik}(x_1,x_2,\mu_A,\mu_B) H_p^{jl}(x_1^\prime,x_2^\prime,\mu_A,\mu_B) \nonumber\\
&& \times d\hat{\sigma}_{ij}(x_1,x_1^\prime,\mu_A) d\hat{\sigma}_{kl}(x_2,x_2^\prime,\mu_B) dx_1 dx_2  dx_1^\prime dx_2^\prime \,,
\label{eq:dps-xs}
\end{eqnarray}
where $m$ is a symmetry factor which is equal to 1 (2) if the two hard-scattering subprocesses are identical (nonidentical).   The {\it joint probabilities} $H_p^{i,k}(x_1,x_2,\mu_A,\mu_B)$  can be approximated as the product of two single PDFs:
\begin{equation}
H_p^{i,k}(x_1,x_2,\mu_A,\mu_B) = f_p^i (x_1,\mu_A) f_p^k(x_2,\mu_B) \,.
\label{eq:joint-PDF}
\end{equation}
Given that one hard scattering has taken place, the parameter $\sigma_{\rm eff}$ measures the size of the partonic core in which the flux of accompanying short-distance partons is confined.  Typical values in phenomenological studies focus on the 10-12 mb region, consistent with measurements from the Tevatron collider~\cite{Abe:1997xk,D0:2009}.  In writing Eqs.~(\ref{eq:dps-xs}) and (\ref{eq:joint-PDF}), we ignore possible strong correlations in longitudinal momentum.  However, for the small values of $x$ expected at the LHC, this should be a good approximation~\cite{Berger:2009cm}.

In order to observe a DPS signal, it is advantageous if: 1) the cross sections for the two individual processes which make up the DPS process are large (in the mb - pb range), and 2) the final state contains objects which can be easily tagged (or identified).  The $b\bar{b}jj$ channel possesses both of these qualities.  By focusing on distributions which contain kinematic information about the {\it entire} final state, we were able to show~\cite{Berger:2009cm} that the DPS component of the final state can be observed despite the presence of a dominant SPS component over a wide kinematic range.  

In this paper, we examine another final state which may be a good candidate to observe DPS, namely the production of a $W$ boson in association with a pair of bottom quark jets.  In the DPS contribution to this final state, one hard scattering produces the $W$ via the Drell-Yan mechanism, while the other hard scattering produces a $b\bar{b}$ pair.  The cross sections for these two processes are individually large, and the charged lepton from the $W$ decay (along with the bottom quarks in the final state) provides a relatively clean signal to tag on.   Our purpose is to establish whether double parton scattering can be 
observed as a discernible physics process in $Wb\bar{b}$ production at LHC energies.  We remark, however, that the $Wb\bar{b}$ final state is a significant background for the production of a Higgs boson $H$ in the $HW^\pm$ mode, where the Higgs boson decays into a pair of bottom quarks \cite{DelFabbro:1999tf}, and that it can be a background in channels where new physics may arise such as in single top quark production~\cite{Cao:2007ea}.  Once DPS production of  $Wb\bar{b}$ is observed, it would be interesting to assess its potential significance as a background in such searches.   A realistic 
study would require knowledge of the effective cross section for double parton scattering in the $Wb \bar{b}$ channel and a set of optimized physics cuts pertinent for the search in question.   We leave this 
topic for possible future work.  

The rest of the paper is structured as follows.  In Sec. \ref{sec:procedure}, we outline our procedure for computing the DPS and SPS contributions to $Wb\bar{b}$ at the LHC, and we discuss and evaluate backgrounds to the same final state.  Our SPS and DPS event rates are computed at next-to-leading order with the aid of the POWHEG BOX code~\cite{POWHEG}.  This section includes a specification of the basic acceptance cuts we use in defining the event sample.  Section \ref{sec:ttbar-bkgd} is devoted to the role of the large $t \bar{t}$ background.   We find that a cut to eliminate events with large missing transverse energy is effective in suppressing this background.   
In Sec. \ref{sec:DPS-vs-SPS}, we focus on discrimination between the DPS and SPS contributions to $Wb\bar{b}$.  We study various single variable and two-dimensional kinematic distributions to bring out the DPS contribution more cleanly.  By utilizing cuts that enhance the DPS $Wb\bar{b}$ sample, we find that the DPS signal can be observed with a statistical significance in the range $S/\sqrt{B} \sim 12 - 15$.  
Section~\ref{sec:conclusions} contains our summary.

\section{DPS and SPS contributions to $Wb\bar{b}$ production at the LHC}
\label{sec:procedure}

We begin with the premise that there are DPS and SPS components of the same $Wb\bar{b}$ final state.  Our aim is to try to pick out the DPS component and to study its distinct properties.  In this section, we outline our method for computing event rates for $Wb\bar{b}$ from DPS and SPS as well as the backgrounds for the same final state at the LHC.  We perform all calculations at a center-of-mass energy of $\sqrt{s} = 7$ TeV.  Event rates are quoted for 10 fb$^{-1}$ of integrated luminosity.  For the DPS case, $Wb\bar{b}$ production is computed using Eq.~(\ref{eq:dps-xs}) where it is assumed that one hard scattering produces the $W$ boson via the Drell-Yan mechanism ($q \bar{q} \rightarrow W^\pm$ at leading order), while the other scattering produces the $b\bar{b}$ system (with either $gg\to b\bar b$ or $q\bar q\to b\bar b$).  Schematically, we can represent the partonic DPS process as:
\begin{equation}
\left( ij \to W^\pm \right) \otimes \left( k l \to b\bar{b} \right) \,.
\label{eq:DPS-process}
\end{equation}
The individual SPS processes which make up the DPS process in Eq.~(\ref{eq:DPS-process}) are generated using the POWHEG BOX event generator~\cite{POWHEG,Alioli:2008gx,Frixione:2007nw} which includes next-to-leading order (NLO) QCD corrections for both, plus shower emission.  The $\otimes$ symbol denotes the combination of one event from each of the $W^\pm$ and $b\bar{b}$ final states.  All events are produced using the two-loop evaluation of $\alpha_s(\mu)$ (where $\mu = M_W$ for the $W$ process and $\mu = \sqrt{m_b^2 + p_T^2}$ for the $b\bar{b}$ production) and CT10 NLO PDFs~\cite{Lai:2010vv}.

In the SPS production of $Wb\bar{b}$, one hard scattering produces the complete final state.  The events from this process are also generated using the POWHEG BOX~\cite{Oleari:2011ey} which implements the NLO calculation of Ref. \cite{Febres Cordero:2006sj}.

Extracting evidence for DPS $Wb\bar{b}$ production is complicated by the fact that many standard model processes imitate the $Wb\bar{b} \to b\bar{b} \ell \nu$ final state.  In particular, we consider 
contributions from the following final sates:
\begin{itemize}
\item Top quark pair production $t \bar{t}$ where either ({\it{i}}) both $t$'s decay semileptonically (denoted by $t_\ell t_\ell$), and one of the charged leptons is missed, or ({\it{ii}}) where one $t$ decays semileptonically while the other decays hadronically (denoted by $t_\ell t_h$) and two jets are either missing or do not pass the threshold and isolation cuts. 

\item Single top quark production ($t\bar{b}$, $\bar{t}b$, $tj$ and $\bar{t}j$ modes) where $t \to W^+ b \, (\bar{t} \to W^- \bar{b})$.
\item $Wjj$, where both light jets are mistagged as a $b$ jets.
\item $Wbj$ where the light jet is mistagged as a $b$ jet.
\end{itemize}
We also considered the following processes, which have a negligible contribution after cuts:
\begin{itemize}
\item $b\bar{b}j$ where one $b$ quark gives an isolated lepton and the light jet is tagged as a $b$ jet.
\item $Zb\bar{b}$ where one lepton from the $Z$ decay goes missing.
\item $b\bar{b}b\bar{b}$ ($b\bar{b}c\bar{c}$) production where at least one heavy quark gives an isolated lepton and the other does not pass the threshold cuts.
\end{itemize}
The $Wjj$ background (where both jets fake bottom quark jets) can be produced in both SPS and DPS processes.  We compute the DPS contribution using the same method discussed above for $Wb\bar{b}$, using the additional POWHEG code~\cite{Alioli:2010xa},  and we include it in our analysis.  The top pair~\cite{Frixione:2007nw} and single top~\cite{Alioli:2009je} SPS processes are also generated using the POWHEG BOX.  Other SPS processes are generated using MadEvent~\cite{Alwall:2007st} or ALPGEN~\cite{Mangano:2002ea}; $Wjj$ is reweighted using a $K$ factor obtained with MCFM~\cite{Campbell:2003hd}.  

In order to avoid soft and collinear divergences in the processes that we compute at LO, we apply a minimal set of {\it generator-level} cuts:
\begin{equation}
p_{T,j} > 15\,{\rm{GeV}} \,\,\, , \,\,\, |\eta_j| < 4.8 \,\,\, , \,\,\,  |\eta_\ell| < 2.5 \,,
\end{equation}
\begin{equation}
p_{T,b} > 15\,{\rm{GeV}} \,\,\, , \,\,\, |\eta_b| < 2.5 \,
\end{equation}
\begin{equation}
\Delta R_{j(b)j(b)} > 0.4 \,\,\, , \,\,\, \Delta R_{j(b)\ell} > 0.4 \,,
\end{equation}
where $\eta$ is the pseudorapidity and $\Delta R_{l k}$ is the separation in the azimuthal-pseudorapidity plane between the two objects $l$ and $k$:
\begin{equation}
\Delta R_{l k} = \sqrt{ \left( \eta_l - \eta_{k} \right)^2 + \left( \phi_l - \phi_{k} \right)^2 } \,.
\end{equation}
Some of these generator-level cuts cannot be applied for processes computed with POWHEG, but they 
are applied subsequently to ensure equal treatment of all event samples.

\subsection{Simulation}

We concentrate on the final state in which there are two $b$ jets, a hard lepton, and missing transverse energy $\met$.   To identify the $Wb\bar{b}$ final state and reduce backgrounds, we begin with simple identification cuts on the generated event samples.  First, we consider only leptonic decays of the $W$ boson ($W \to \ell \nu$).  We focus on the case $\ell = \mu$, since electrons with low transverse momentum can be easily faked by light jets.  We limit the hadronic activity in our events to include exactly two hard jets, both of which must be identified as bottom quark jets.  Finally, all events (DPS and SPS $Wb\bar{b}$ as well as backgrounds) are required to pass the following {\em acceptance} cuts:
\begin{equation}
p_{T,b} \ge 20 \,{\rm{GeV}} \,,\, |\eta_b| \le 2.5 \,,
\end{equation}
\begin{equation}
20 \, {\rm{GeV}} \le p_{T,\mu} \le 50 \, {\rm{GeV}} \,, \, |\eta_\mu| < 2.1 \,,
\end{equation}
\begin{equation}
\met \ge 20 \, \rm{GeV}
\end{equation}
\begin{equation}
\Delta R_{b\bar{b}} \ge 0.4 \,,\, \Delta R_{b\mu} \ge 0.4 \,.
\end{equation}

The cut on the missing transverse energy $\met \ge 20 \, \rm{GeV}$ is motivated by the fact that the 
neutrino momentum in $W$ decay is not observed.   The $20$~GeV cut on the $b$ jets and the 
lepton is invoked to eliminate contributions from the underlying event.  The upper lepton $p_T$ cut is used to reject boosted $W$ bosons, as in the case where a $W$ boson originates from a $t$-quark decay, or when the $W$ recoils against the $b\bar b$ pair in SPS.
  
To account for $b$ jet tagging efficiencies, we assume a $b$-tagging rate of 60\% for $b$ quarks with $p_{T,b} > 20\text{ GeV}$ and $|\eta_{b}| < 2.5$.  We apply the muon identification efficiencies found in the ATLAS Technical Design Report~\cite{ATLASTDR}.
Detector resolution effects are accounted for by smearing the final-state energy according to:
\begin{equation}
\frac{\delta E}{E} = \frac{a}{\sqrt{E/{\rm{GeV}}}} \oplus b \,,
\end{equation}
where $a = 50\%$ and $b = 3\%$ for jets and $a = 10\%$ and $b = 0.7\%$ for leptons.  Light jets (jets from $u,d,s$ and $c$ quarks as well as gluons) can ``fake" bottom quark jets and we account for this by applying a mistagging rate for the gluon and $u,d$ and $s$ quarks of:
\begin{equation}
\epsilon_{u,d,s,g \to b} = 0.67\% 
\end{equation}
for $p_{T,j} < 100$ GeV and:
\begin{equation}
\epsilon_{u,d,s,g \to b} = 2\% 
\end{equation}
for $p_{T,j} > 250$ GeV.  For $p_T$ values between 100 and 250 GeV, we linearly interpolate the fake rates.  Finally, for $c$ quarks, we assume a fake rate of:
\begin{equation}
\epsilon_{c \to b} = 10\% 
\end{equation}
for $p_{T,c} > 50$ GeV and we linearly interpolate fake rates for $p_T < 50$ GeV.

Table \ref{tbl:cut-results} shows the number of events from the $Wb\bar{b}$ final state (DPS and SPS) and the backgrounds both before (column labeled ``generator-level cuts") and after the acceptance cuts, detector effects, and mistagging effects are applied (column labeled ``acceptance cuts").   In these results and those that follow, we sum the $W^+$ and $W^-$ events.   In evaluating the DPS processes, we assume a value $\sigma_{\rm eff} \simeq 12$ mb for the effective cross section.  However, we stress that the goal is to motivate an empirical determination of its value at LHC energies.  The acceptance cuts are very effective against the $W j j$ final states, both for DPS and SPS.   The results in Table \ref{tbl:cut-results}  make it  apparent that $Wb\bar{b}$ production from SPS and the top quark pair background are the most formidable obstacles in extracting a DPS signal.    We address background rejection in the next two sections.  

\begin{table}[t]
\begin{center}
\begin{tabular}{c|c|c|c|c}
\hline
\,\,\, Process \,\,\,& \,\,\, Generator-level cuts \,\,\, & \,\,\, Acceptance cuts \,\,\,& \,\,\, $\met \le 45$ GeV \,\,\,& \,\,\, $S_{p_T}^\prime \le 0.2$ \,\,\, \tabularnewline
\hline
\hline
$W^\pm b \bar{b}$ (DPS) & 10000 & 247 & 231 & 173 \tabularnewline
\hline
\hline
$W^\pm b \bar{b}$ (SPS) & 44000 & 1142 & 569 & 114 \tabularnewline
\hline
\hline
$t\bar{t}$ & 225000 & 1428 & 290 & 13 \tabularnewline
\hline
$W^\pm jj$ (DPS)  & 476000 & 43.5 & 37.7 & 27.3 \tabularnewline
\hline
$W^\pm jj$ (SPS)  & 20300000 & 101 & 55.7 & 19.6 \tabularnewline
\hline
Single top & 20000 & 492 & 168 &  15 \tabularnewline
\hline
$W^\pm b j$ & 153000 & 152 & 53.1 & 8.2 \tabularnewline
\hline
\hline
\end{tabular}
\end{center}
\caption{Numbers of events before and after the various cuts are applied for 10 fb$^{-1}$ of data.  After acceptance cuts, SPS $Wb\bar{b}$ production and $t\bar{t}$ production dominate the event rate.   A maximum $\met$ cut reduces the background from $t\bar{t}$ significantly.  A maximum cut on $S_{p_T}^\prime$ improves the DPS/SPS ratio in $Wb\bar{b}$ production.}
\label{tbl:cut-results}
\end{table}

\
\section{$t\bar{t}$ Background Rejection}
\label{sec:ttbar-bkgd}

\begin{figure}[h!]
\includegraphics[scale=0.5]{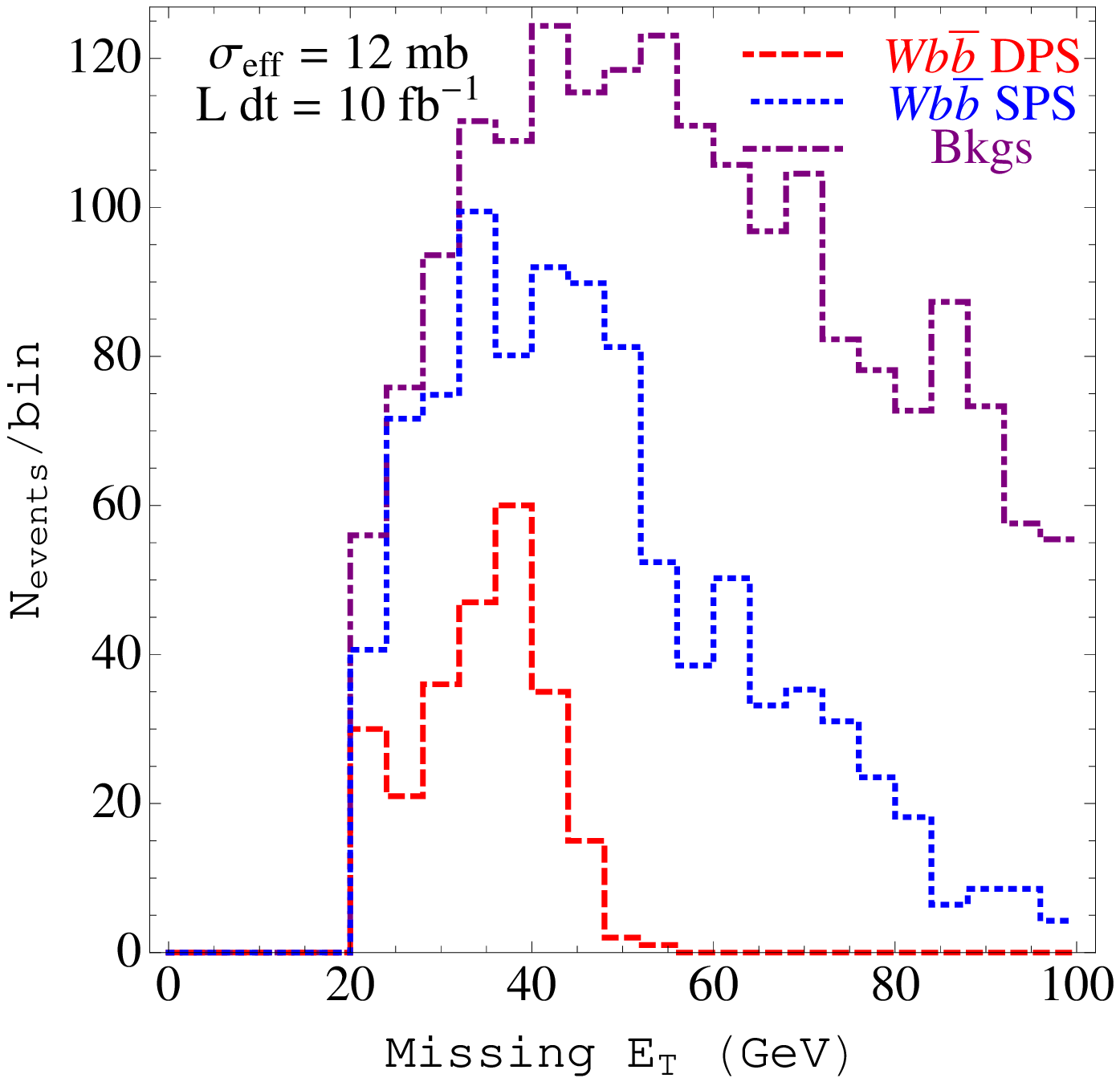}
\hspace{1cm}
\includegraphics[scale=0.5]{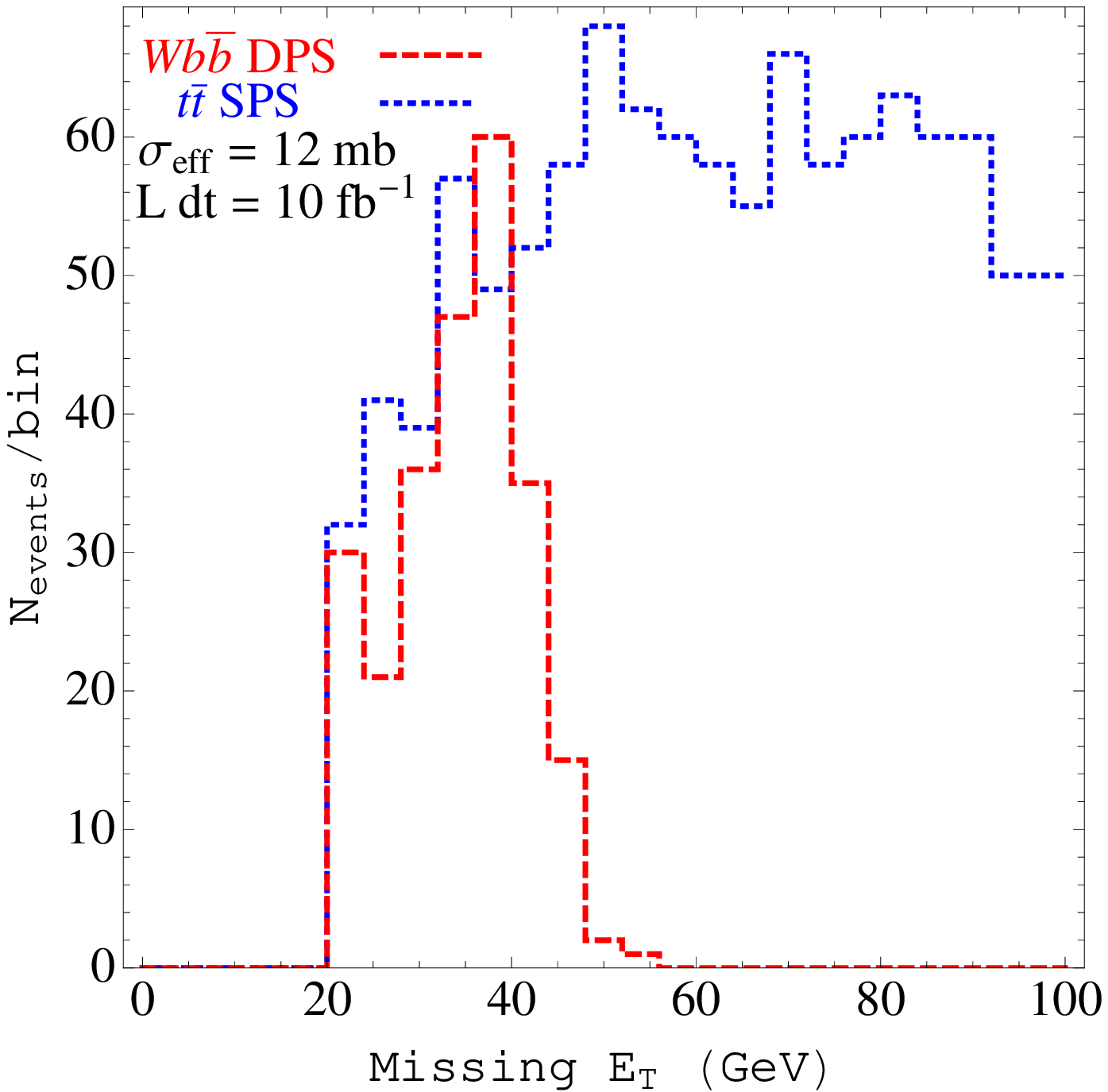}
\caption{The event rate as a function of $\met$ for DPS and SPS.  On the left, all backgrounds are included while the plot on the right compares the DPS events to those from $t\bar{t}$ alone.  While the DPS signal is concentrated in the $\met < 45$ GeV range, the majority of the $t\bar{t}$ background lies above this range.  Therefore, imposing an maximum $\met$ cut of 45 GeV can greatly reduce the background coming from $t\bar{t}$ production. }
\label{fg:ET-miss}
\end{figure}

We examine three possibilities to reduce the $t\bar{t}$ background:  a cut to restrict $\met$ from above, 
rejection of events in which a top quark mass can be reconstructed, and a cut to restrict the transverse momentum of the leading jet.    In the end, an upper cut on $\met$ in the event appears to offer the best advantage.  Indeed, one would expect that $\met$ in $Wb\bar{b}$ events would be smaller than $\met$ in $t\bar{t}$ events.  Top quark decays give rise to boosted $W^\pm$'s which, after decay, should result in larger values of missing $E_T$ compared to the $Wb\bar{b}$ process.   The $\met$ distribution is shown in Fig.~\ref{fg:ET-miss} for the DPS component of $Wb\bar{b}$, the SPS component of $Wb\bar{b}$, and {\it{all}} backgrounds (left).  On the right, we show the DPS component of $Wb\bar{b}$ and the $t\bar{t}$ background alone.  The plot on the right shows that the DPS signal is produced in the region of relatively small $\met$ and the $t\bar{t}$ background has a harder spectrum in $\met$.   One way to suppress the $t\bar{t}$ background while leaving the DPS signal unaffected is to impose a maximum $\met$ cut in the 40-60 GeV range.  In the analysis that follows, we include a maximum $\met$ cut of 45 GeV in addition to the acceptance cuts outlined above.  
     
The effects of the maximum $\met$ cut are shown in the fourth column of Table~\ref{tbl:cut-results}.   
This cut eliminates about 80\% of the $t\bar{t}$ background that remained after the initial acceptance cuts.  The cut is also effective at reducing the single top quark and $W b j$ backgrounds, 
eliminating about 67\% in both cases.  On the other hand, 93\% of the DPS  $Wb\bar{b}$ events and 
46\% of the SPS $Wb\bar{b}$ events are retained.  

Backgrounds from events that contain a real top quark, such as the $t\bar{t}$ and single top events, might be reduced if one could reconstruct a top quark mass distribution $M_{b\ell\nu}$ 
from the final-state objects (bottom quarks, charged leptons, and neutrinos), and then eliminate events in 
which the reconstructed mass falls in a narrow window centered on the known top quark mass.  To accomplish this, we must know the value of the longitudinal momentum for the neutrino.  We compute this momentum via the on-shell mass relations of the $W$-boson decay.  The quadratic nature of the mass relations produces a twofold ambiguity.  In our analysis, we include both solutions for the neutrino momentum.  After reconstruction, events that result in a value of $M_{b\ell\nu}$ within a window (of 10, 15, or 20 GeV) around the measured top quark mass (we assume $m_t = 175$ GeV) are rejected.  For all three values of the window, we find a slight improvement of signal-to-background ratio but accompanied by a decrease in the significance ($S/\sqrt{B}$) of the signal associated with the overall decrease in event rate.  

The poor discriminating power of this mass reconstruction method results from two issues.  First, the 
twofold ambiguity for the longitudinal component of the neutrino momentum provides a combinatorial background.  In addition, while the $t\bar t$ events which pass the acceptance cuts are predominantly from the $t_\ell t_h$ decay mode, about 30\% are from the $t_\ell t_\ell$ mode.  With two neutrinos in the final state, the on-shell mass relations are not applicable and do not provide a unique set of neutrino momenta.  For these reasons, the mass reconstruction observable is not considered a good discriminator.  

\begin{figure}[h!]
\includegraphics[scale=0.5]{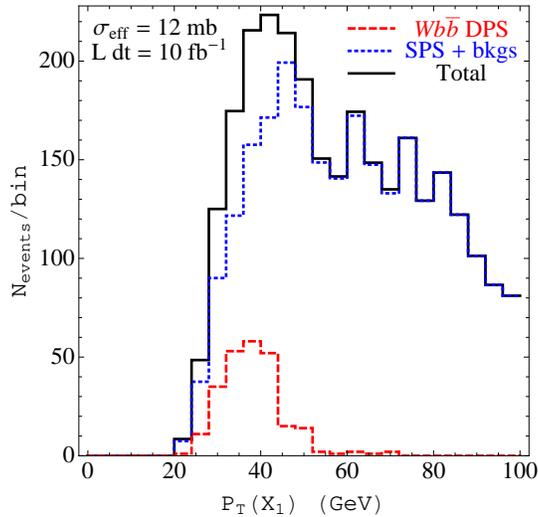}
\caption{The event rate as a function of the transverse momentum of the leading object (either a jet or lepton).  The SPS contribution has a harder $p_T$ spectrum.}
\label{fg:pt-leading}
\end{figure}

Jets from final states that contain top quarks tend to have a hard spectrum, associated with the large top quark mass.   A possible observable for $t\bar{t}$ background rejection is therefore the transverse momentum of the leading object (either a jet or a charged lepton).   The $p_T$ spectrum of the leading object tends to be soft in DPS events~\cite{Berger:2009cm}.   However, SPS production of 
$Wb\bar{b}$ yields a rather hard $p_T$ spectrum since the bottom quarks recoil against the $W^\pm$ boson.  In Fig.~\ref{fg:pt-leading}, we show the $p_T$ distributions for the leading object in $Wb\bar{b}$ DPS production and for the remainder (which includes SPS $Wb\bar{b}$ production).  We see that the DPS events indeed populate the lower bins of the allowed $p_T$ spectrum, while the SPS and background events result in a harder spectrum.   When we compare the usefulness of placing a cut on the  upper value of $p_T$ with the improvement we find with the cut on the maximum value of $\met$, 
we conclude that the maximum $\met$ cut provides better significance.  If we use both cuts, we find that the DPS signal significance is degraded.  

We conclude this section with the statement that of the three possibilities to reduce the $t\bar{t}$ background we considered, a cut to restrict $\met$ from above appears to offer the best advantage, and it is the only cut we impose in addition to the acceptance cuts specified above.

\section{Discrimination between DPS and SPS contributions to $Wb\bar{b}$ }
\label{sec:DPS-vs-SPS}

To separate the DPS events from those of SPS origin, we find it convenient to employ quantities which take into account information from the entire final state.  One useful observable is $S^\prime_{p_T}$, defined as \cite{D0:2009}:
\begin{equation}
S^\prime_{p_T} = \frac{1}{\sqrt{2}} \sqrt{ \left( \frac{|p_T(b_1,b_2)|}{|p_T(b_1)| + |p_T(b_2)|}\right)^2 + \left( \frac{|p_T(\ell,\met)|}{|p_T(\ell)| + |\met|}\right)^2 } \,.
\label{eq:Sptprime}
\end{equation}
In our case $p_T(b_1,b_2)$ is the vector sum of the transverse momenta of the two $b$ jets, and 
$p_T(\ell,\met)$ is the vector sum of $\met$ and the transverse momentum of the charged lepton in the final state.  
\begin{figure}[h!]
\includegraphics[scale=0.5]{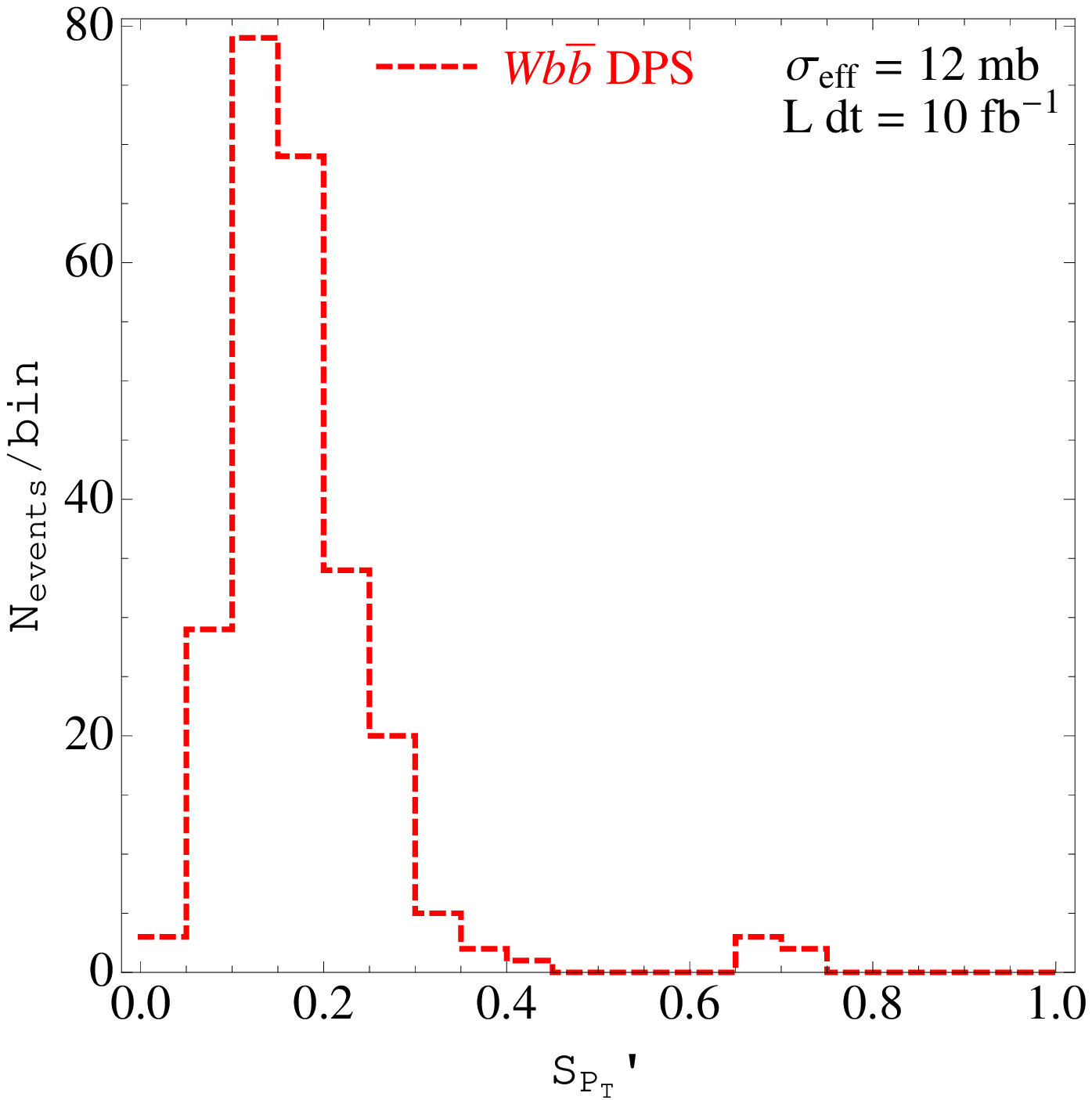}
\hspace{1cm}
\includegraphics[scale=0.5]{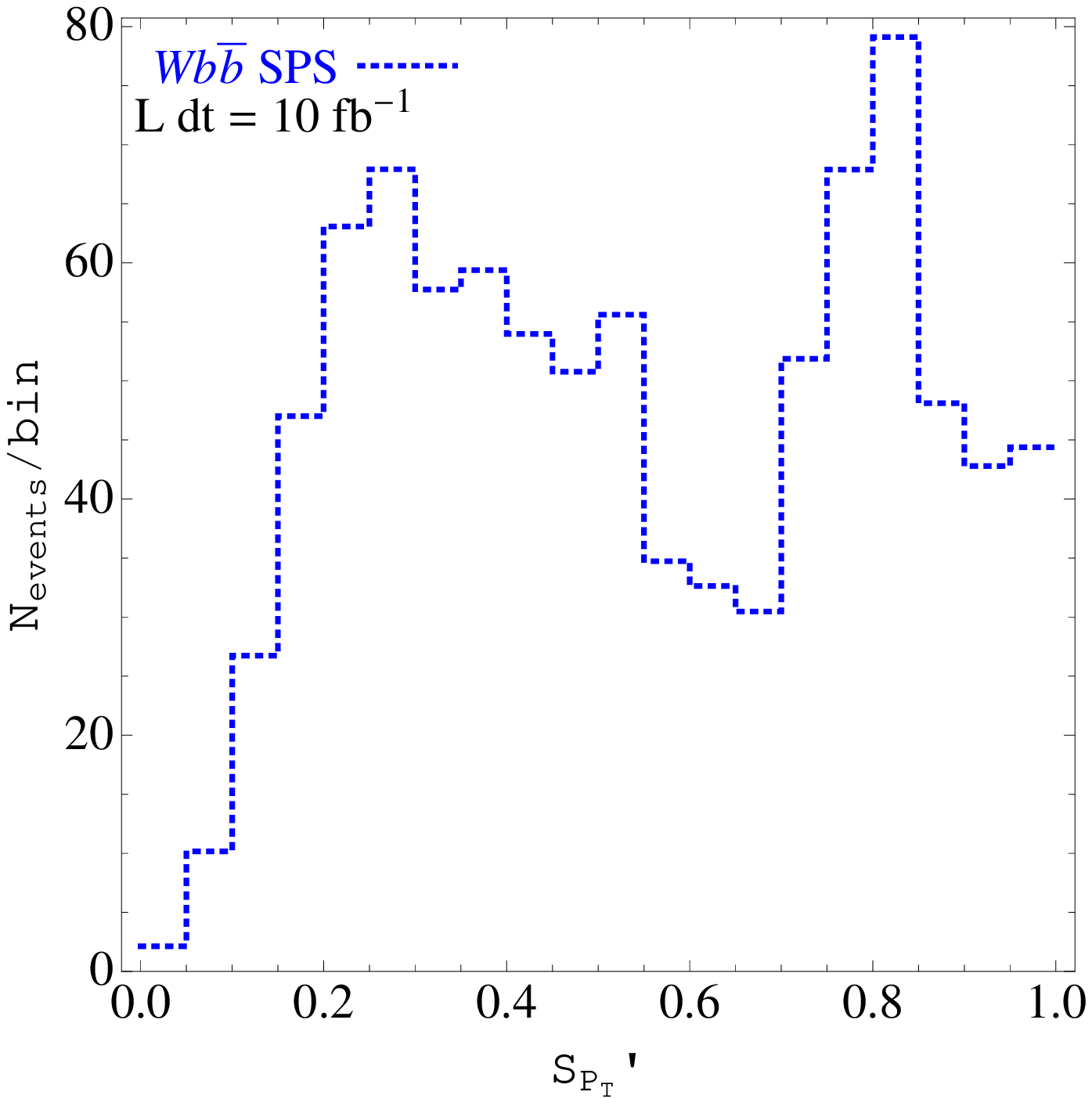}
\caption{The event rate for $Wb\bar{b}$ production from DPS (left) and SPS (right) as a function of $S_{p_T}^\prime$.  In the DPS case, the distribution is peaked toward $S_{p_T}^\prime \simeq 0$; SPS production of $Wb\bar{b}$ produces bottom quarks that are not back-to-back, resulting in a broad distribution and a peak near $S_{p_T}^\prime \simeq 1$.}
\label{fg:sptprime-Wbb-only}
\end{figure} 

In DPS production, the bottom quarks are produced roughly back-to-back such that the vector sum of their transverse momenta tends to vanish.  Likewise, the vector sum of the lepton and neutrino momenta tends to be small (with corrections from the boosted $W^\pm$).  Thus, the $S^\prime_{p_T}$ distribution for the DPS process exhibits an enhancement  at low $S^\prime_{p_T}$, as shown in Fig.~\ref{fg:sptprime-Wbb-only}.  The peak does not occur at exactly $S_{p_T}^\prime = 0$ owing to NLO  real radiation that alters the back-to-back nature of the $b\bar{b}$ and $\ell\nu$ systems.  On the other hand, SPS production of $Wb\bar{b}$ final states does not favor back-to-back configurations, and it exhibits a peak near $S^\prime_{p_T}$ = 1.  This feature is linked to the fact that many $b\bar{b}$ pairs are produced from gluon splitting \cite{Berger:2009cm}.

\begin{figure}[h!]
\includegraphics[scale=0.5]{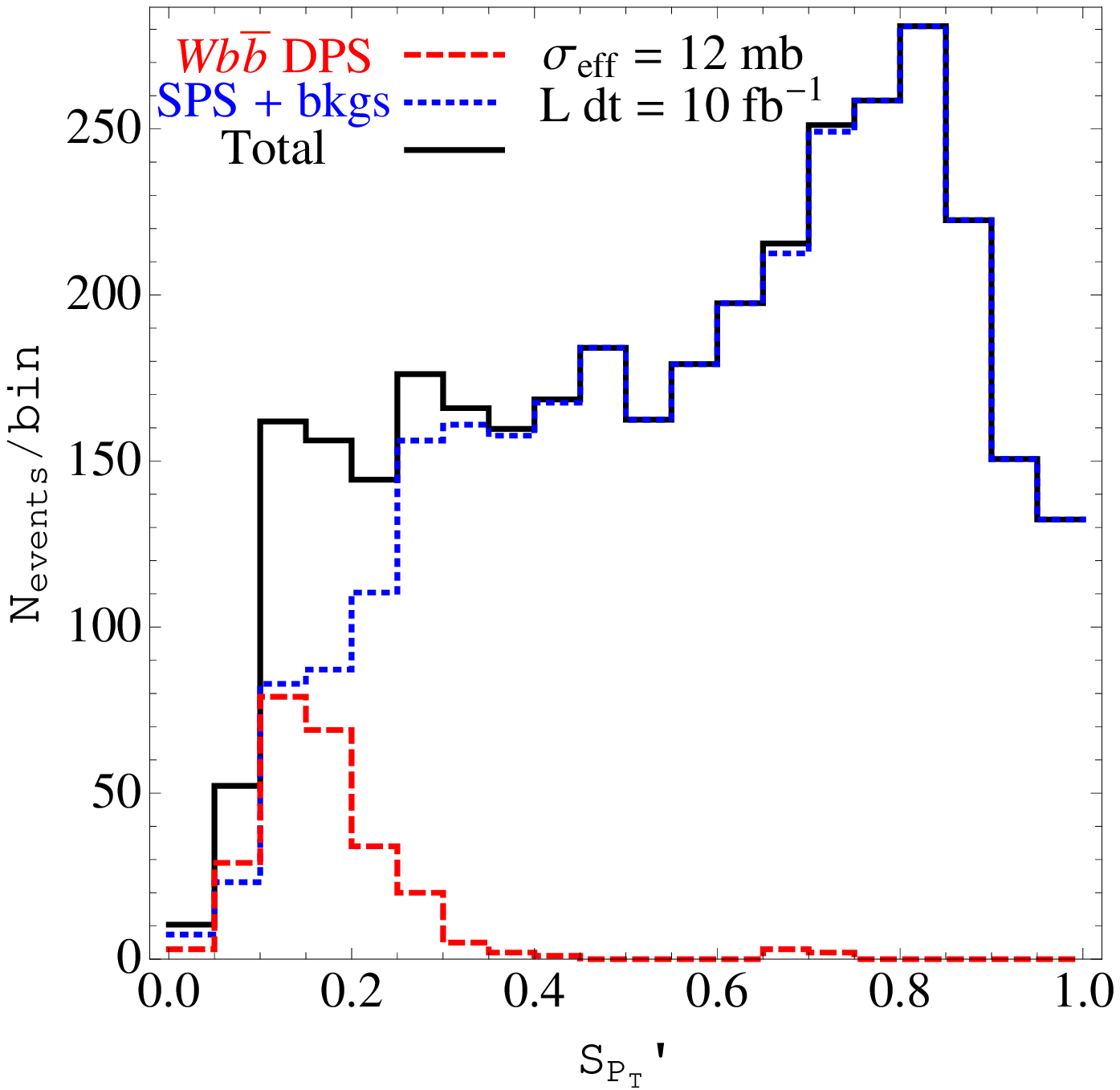}
\hspace{1cm}
\includegraphics[scale=0.5]{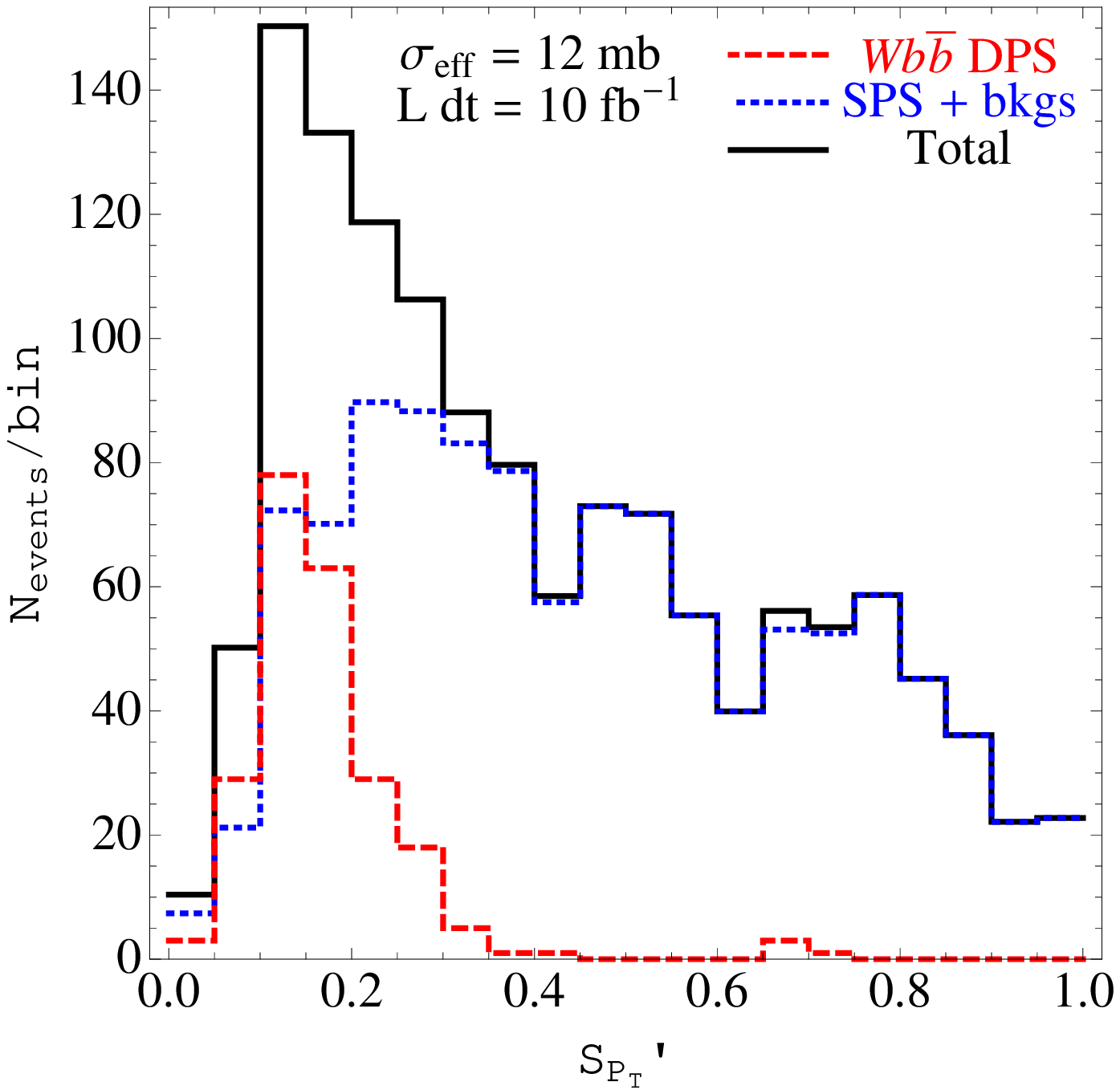}
\caption{The $S_{p_T}^\prime$ distribution for DPS and SPS production of $Wb\bar{b}$ including all relevant backgrounds.  On the left, only the minimal acceptance cuts are imposed, while, on the right, an additional maximum $\met$ cut is imposed ($\met < 45$ GeV).  Imposing a maximum $\met$ cut greatly reduces the background and produces a sharp peak in an $S_{p_T}^\prime$ region where DPS is expected to dominate.}
\label{fg:sptprime-w-and-wo-ETcut}
\end{figure}

The clean separation in $S_{p_T}^\prime$ between the DPS and SPS $Wb\bar{b}$ processes exhibited in Fig.~\ref{fg:sptprime-Wbb-only} is obscured once the $t\bar{t}$ background is included (e.g., see the left side of Fig.~\ref{fg:sptprime-w-and-wo-ETcut}).  Figure~\ref{fg:sptprime-w-and-wo-ETcut} illustrates the effectiveness of the maximum $\met$ cut in reducing the $t\bar{t}$ background in the $S_{p_T}^\prime$ distribution.  After the cut, a sharp peak is evident in the region of small $S_{p_T}^\prime$ where DPS events are expected to reside.   

The plot on the right side of Fig.~\ref{fg:sptprime-w-and-wo-ETcut}, shows that extraction of a relatively clean DPS sample can be accomplished by imposing a maximum $S_{p_T}^\prime$ cut.  The last column of Table~\ref{tbl:cut-results} shows that a cut $S_{p_T}^\prime < 0.2$ reduces the SPS $Wb\bar{b}$ rate while leaving the DPS signal relatively unaffected.  In the end, the major background arises from DPS $Wjj$, as is expected since this process inhabits the same kinematic regions as the DPS $W b\bar{b}$ signal.  Despite this background, we find a statistical significance for the presence of DPS $W b \bar{b}$ of $S/\sqrt{B} = 173/\sqrt{197} = 12.3$.

\subsection{Further discrimination}
\label{subsec:other-bkgds}

\begin{figure}[h!]
\includegraphics[scale=0.5]{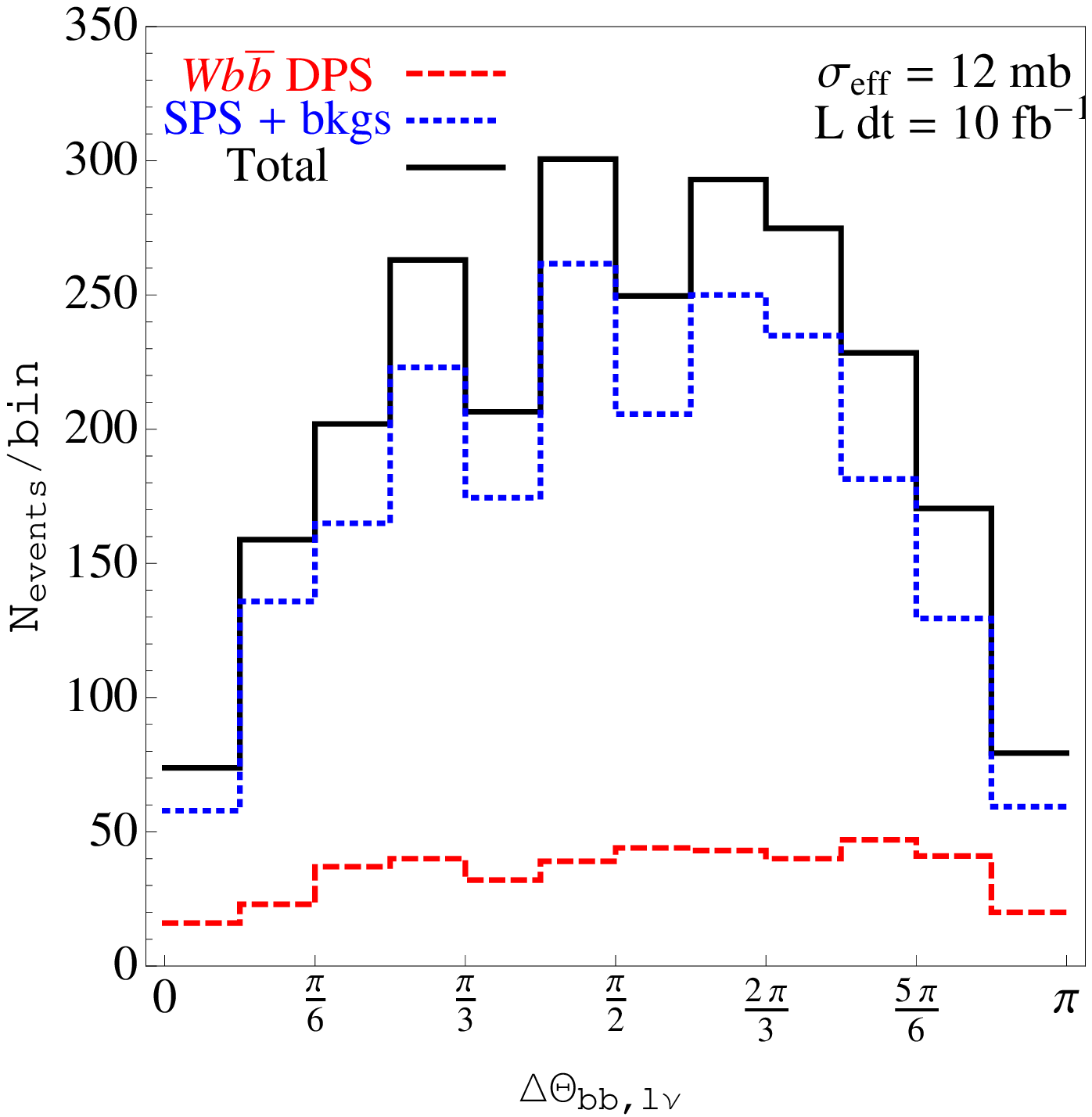}
\hspace{1cm}
\includegraphics[scale=0.5]{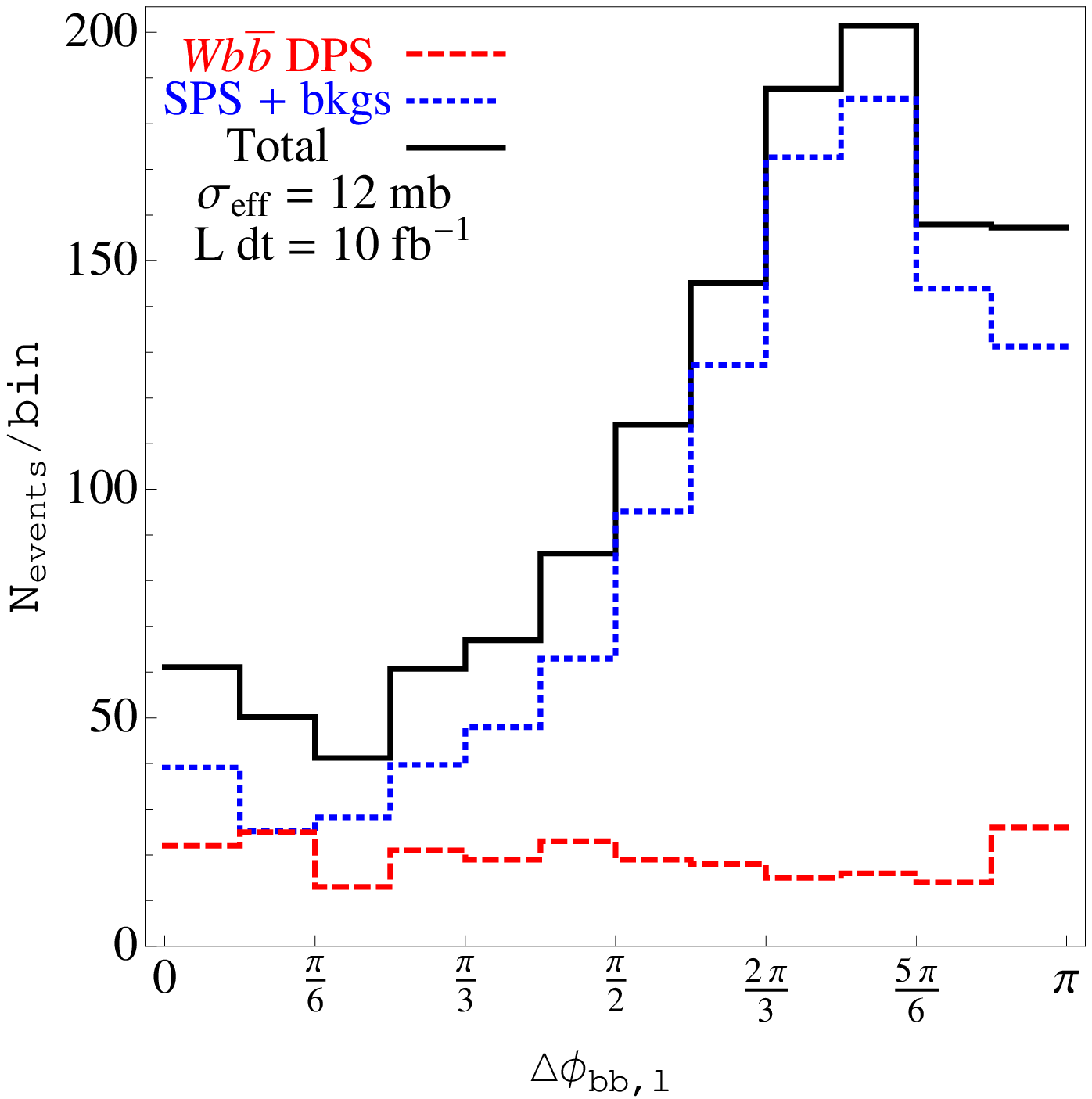}
\includegraphics[scale=0.5]{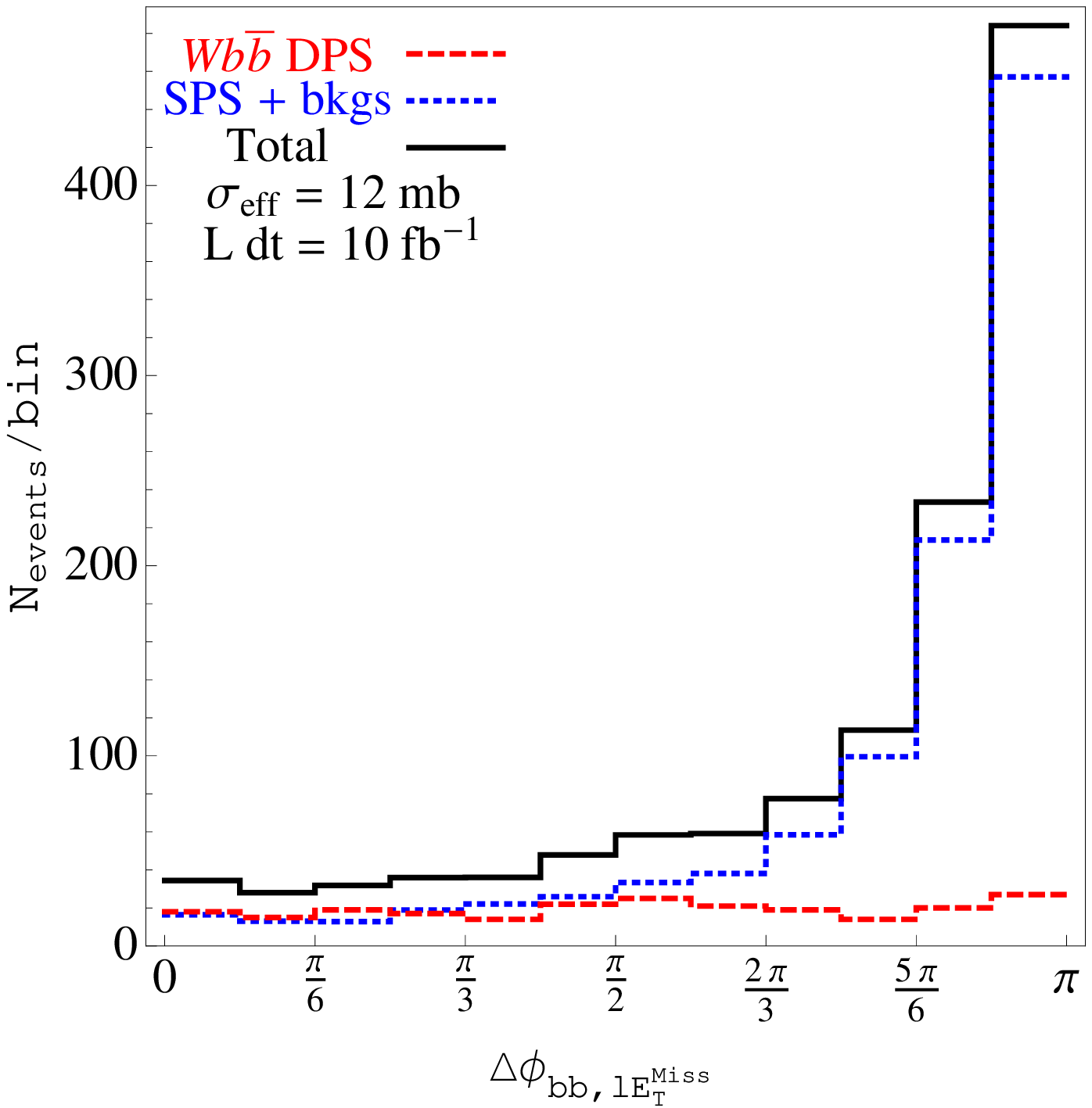}
\caption{The event rate as a function of the angle between the normals of the two planes defined by the $b\bar{b}$ and $\ell \nu$ systems (top-left), the azimuthal angle between the charged lepton and the total momentum vector of the  $b\bar{b}$ system (top-right) and the azimuthal angle between the transverse momentum vectors of the $b\bar{b}$ and $\ell\met$ systems (bottom).  In SPS events, it is apparent that there is a strong correlation in the angles.  However, there is no such correlation present in the DPS events.}
\label{fg:delta-phi}
\end{figure} 

Observables which take into account the angular distribution of events are also useful in the search for DPS.  Figure~\ref{fg:delta-phi} depicts three such observables.  In the top-left plot, we show the event rates for DPS $Wb\bar{b}$ and the backgrounds (SPS $Wb\bar{b}$ included) as a function of  the angle between the normals to the two planes defined by the $b\bar{b}$ and $\ell\nu$ systems.  These planes are defined in the partonic center-of-mass frame and are specified by the three-momenta of the outgoing jets or leptons.  The angle between the two planes defined by the $b\bar{b}$ and $\ell\nu$ systems is:
\begin{equation}
\cos \Delta\Theta_{b \bar{b},\ell\nu} = \hat{n}_3(b_1, b_2) \cdot  \hat{n}_3(\ell,\nu)
\label{eq:cosPhi-def}
\end{equation}
where $\hat{n}_3(i,j)$ is the unit three-vector normal to the plane defined by the $i-j$ system and $b_1 (b_2)$ is the leading (next-to-leading) $b$ jet.  In order to construct the normals $\hat{n}_3(b_1,b_2)$ and $\hat{n}_3(\ell,\nu)$, we require full event reconstruction using the on-shell $W$-boson mass relations.  We see that the distribution of the DPS events is rather flat, aside from the
cut-induced suppressions at $\Theta_{b\bar b, \ell \nu} \sim 0$ and $\sim \pi$,  whereas the SPS events show a strong correlation, with a distribution that peaks near $\Delta \Theta_{b\bar{b},\ell\nu} \sim \frac{\pi}{2}$. 

In the top-right plot of Fig.~\ref{fg:delta-phi}, we show the event rates as a function of the azimuthal angle between the charged lepton and the total momentum vector of the $b\bar{b}$ system.  No information from the neutrino is used.  In the bottom plot, we show the event rates as a function of the azimuthal angle between the transverse momentum vectors of the $b\bar{b}$ and $\ell\met$ systems.  Since this azimuthal angle is defined in the transverse plane, it requires only $\met$.  Full event reconstruction to determine the neutrino momentum is not needed.  In both cases, the shape of the DPS distribution is flat while the SPS distribution shows a strong correlation, with a preference for values toward $\pi$. 

In all three plots of Fig.~\ref{fg:delta-phi}, it is clear that DPS and SPS exhibit different behaviors as a function of angular observables.  However, the dominance of SPS $W b \bar{b}$ and backgrounds over DPS $W b \bar{b}$ for the full range of these observables makes it impossible to extract a DPS 
$W b \bar{b}$ signal from these distributions by themselves.

\subsection{Two-dimensional distributions}
\label{subsec:2d-plots}

\begin{figure}[h!].
\includegraphics[scale=0.5]{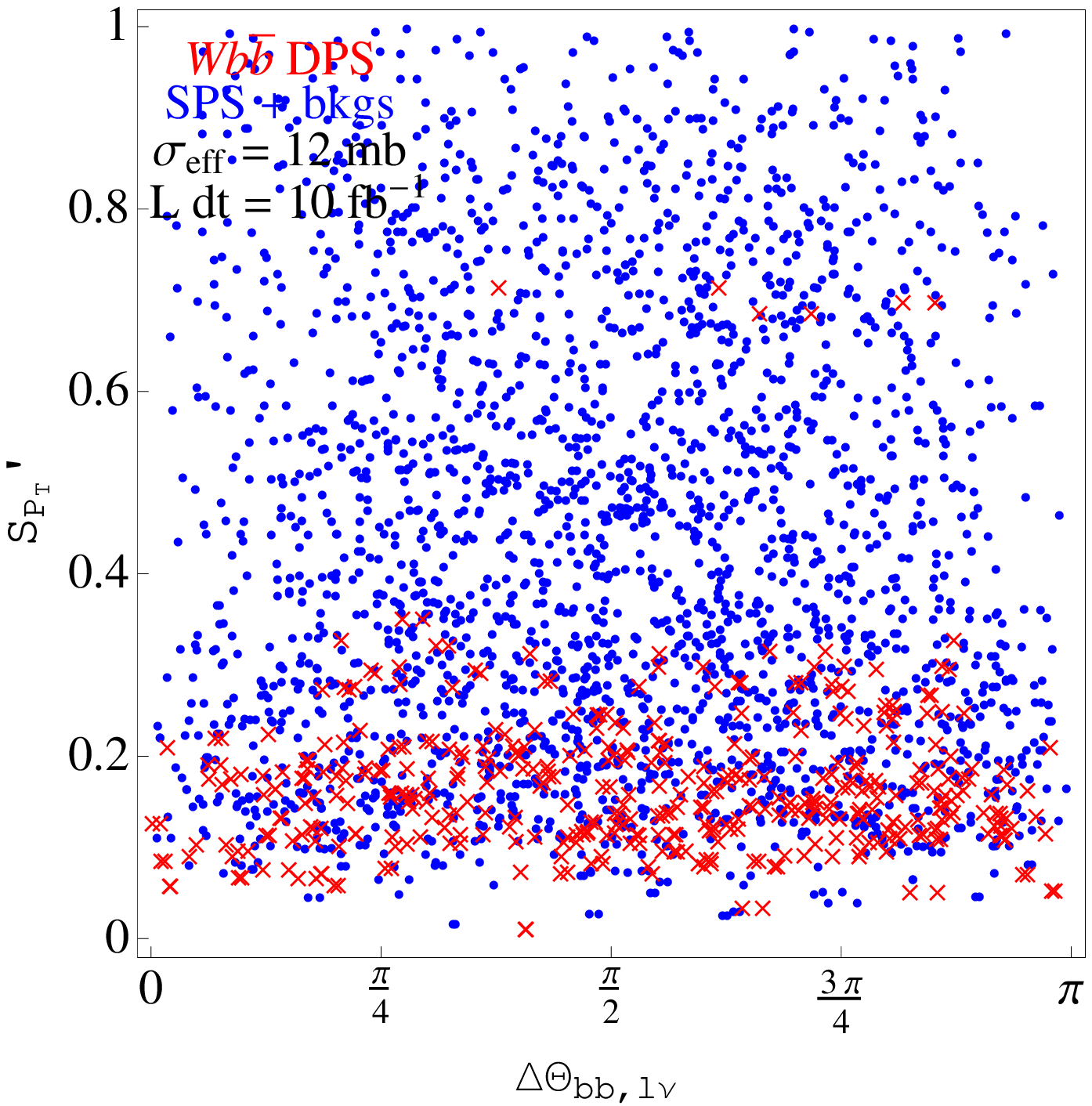}
\hspace{1cm}
\includegraphics[scale=0.5]{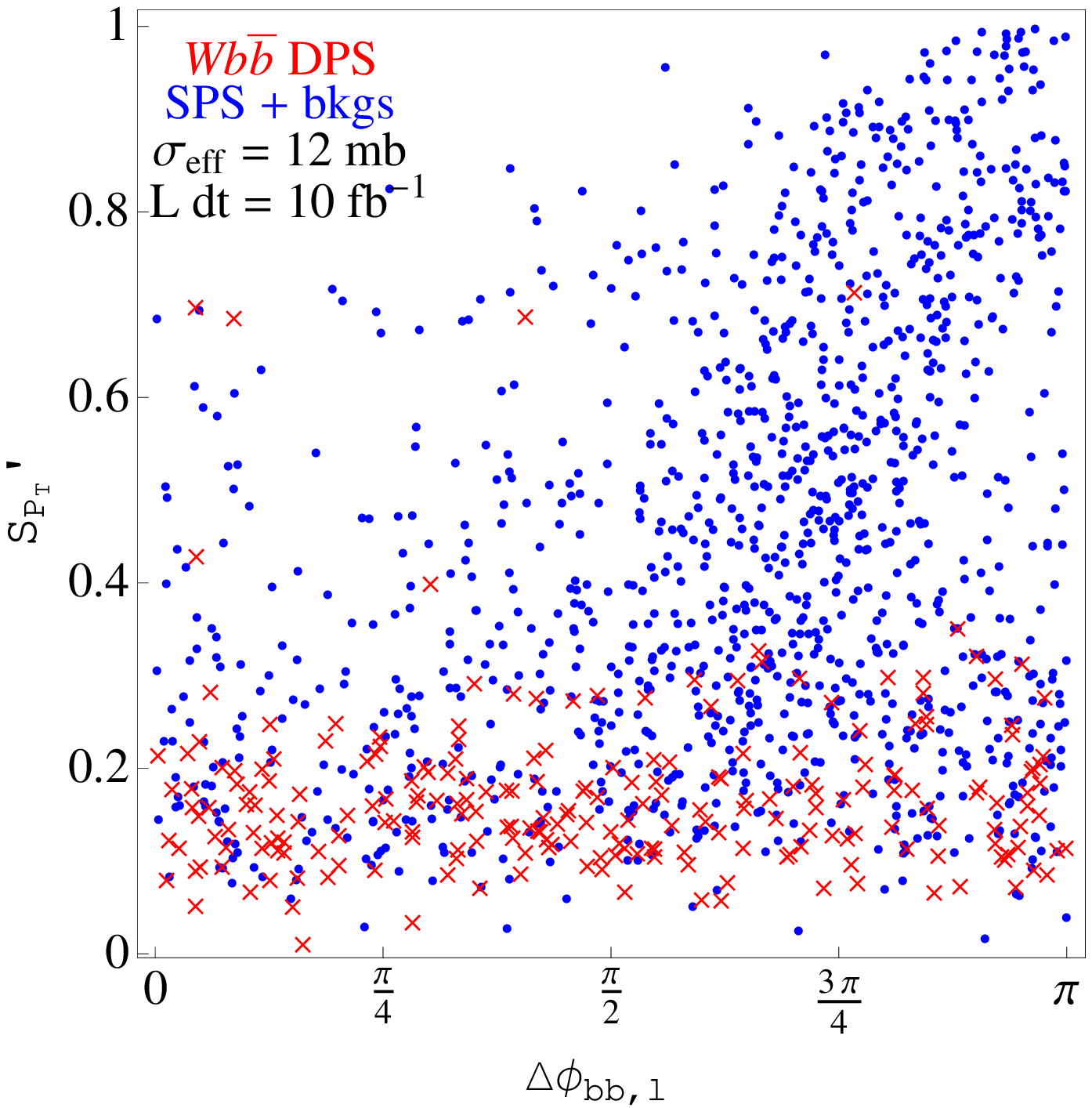}
\caption{Two-dimensional distributions of events in the variables $S_{p_T}^\prime$ and $\Delta \Theta_{b\bar{b},\ell\nu}$ (left) and $S_{p_T}^\prime$ and $\Delta \phi_{b\bar{b},\ell}$ (right).  In both cases, the $W b \bar{b}$ DPS events (denoted by red {\bf{x}}) lie in the lower half of the plane, while the $W b \bar {b}$ SPS and background events (denoted by blue {\bf{dots}}) occupy the upper half.   The plot on the right, in which reconstruction of only the lepton direction is required, appears to achieve a cleaner separation, with SPS and background events concentrated in the upper right-hand corner of the plane.}
\label{fg:scatter-plots} 
\end{figure} 

Despite the dominance of the $W b \bar{b}$ SPS contribution and the backgrounds over the 
DPS $W b \bar{b}$ contribution, the angular distributions can still be extremely useful when used in conjunction with other observables.  Two-dimensional distributions of one variable against another show distinct regions of DPS dominance (or SPS and background dominance).  In Fig.~\ref{fg:scatter-plots}, we construct two such scatter plots.   On the left, we show $S_{p_T}^\prime$ versus the angle between the normals of the two planes defined by the $b\bar{b}$ and $\ell\nu$ systems ($\Delta \Theta_{b\bar{b},\ell\nu}$), while, on the right, we show $S_{p_T}^\prime$ versus the azimuthal angle between the charged lepton and the total momentum vector of the $b\bar{b}$ system.  In both plots, we see that the DPS events reside predominantly in the lower half of the plane (small $S_{p_T}^\prime$) and are distributed evenly in the angular variable.  The separation between DPS $W b \bar{b}$ and the SPS 
component is not as pronounced in the $S_{p_T}^\prime - \Delta \Theta_{b\bar{b},\ell\nu}$ plane as we 
saw in our earlier study of $b \bar{b} j j$~\cite{Berger:2009cm}.  In the $W b \bar{b}$ case, the background events are more evenly distributed over the full plane, to some extent resulting from inclusion of both solutions for the neutrino's longitudinal momentum in the $W^\pm$ decay.  
(The greater density of points in the left plot of Fig.~\ref{fg:scatter-plots} relative to the right plot is 
explained by the fact that both solutions for the neutrino momentum are included in the left plot).  

As shown in the plot on the right of Fig.~\ref{fg:scatter-plots}, the SPS $W b \bar{b}$ and background events in the $S_{p_T}^\prime - \Delta \phi_{b\bar{b},\ell}$ show a strong preference for upper right-hand corner 
of the plane.  This two-dimensional distribution indicates that cuts on the $S_{p_T}^\prime$ and $\Delta \phi_{b\bar{b},\ell}$ variables should permit extraction of an enriched sample of DPS $W b \bar{b}$ events.
\begin{figure}[h!]
\includegraphics[scale=0.5]{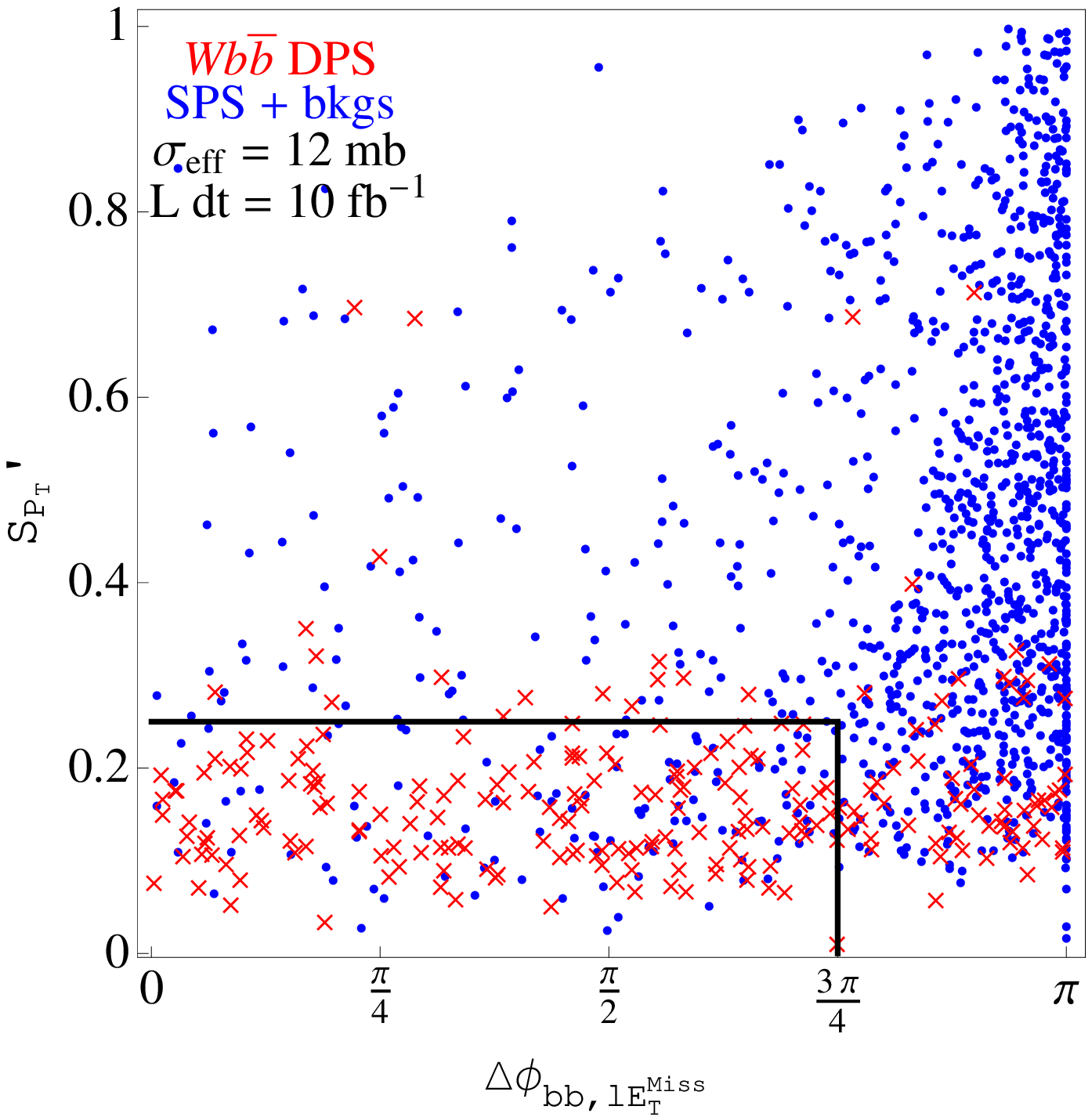}
\hspace{1cm}
\includegraphics[scale=0.5]{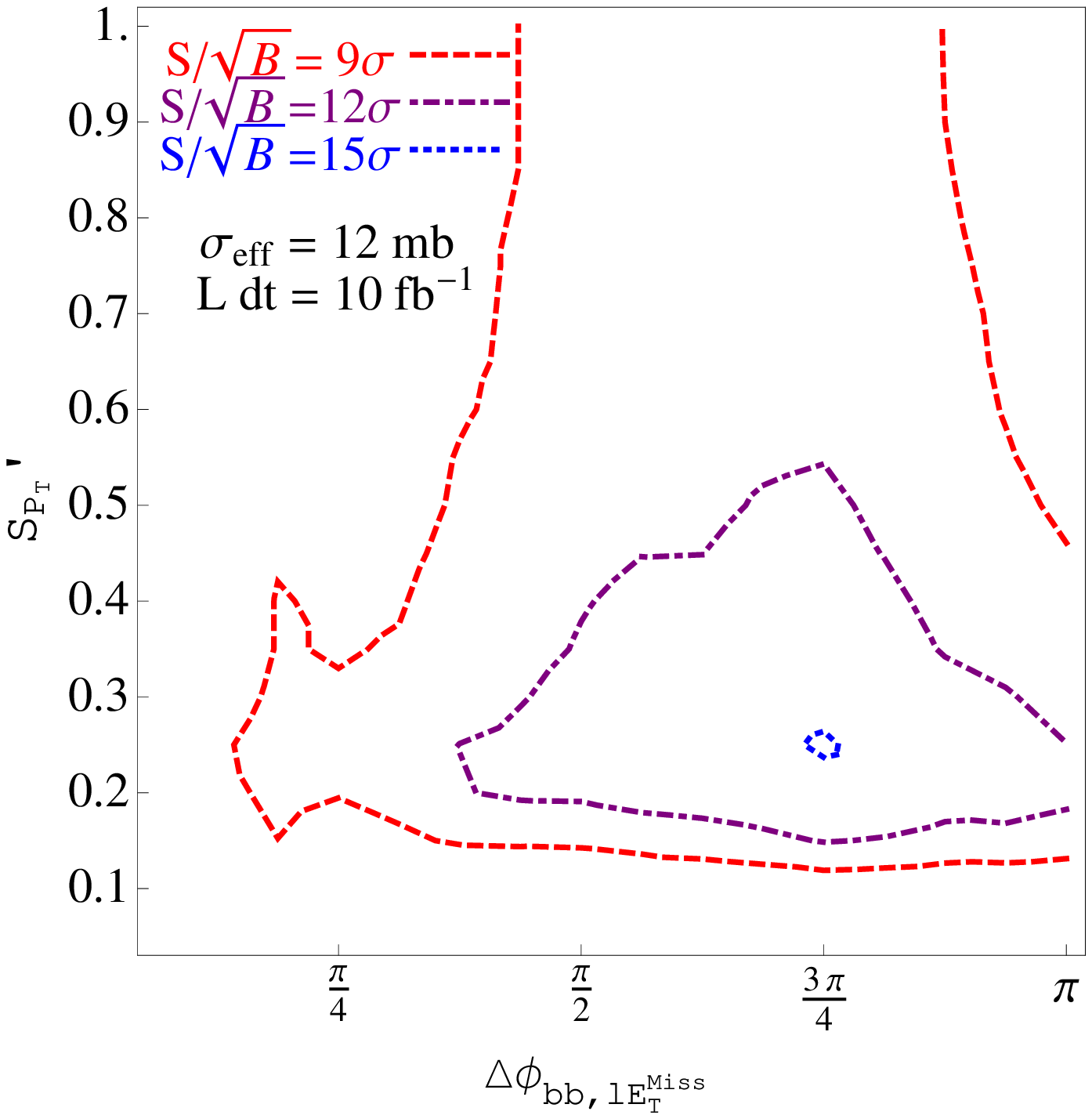}
\caption{The two-dimensional distribution of events in the variables $S_{p_T}^\prime$ and $\Delta \phi_{bb,\ell~\met}$ (left).  The $W b \bar{b}$ DPS events are denoted by red {\bf{x}}, while the $W b \bar {b}$ SPS and background events are denoted by blue {\bf{dots}}.   The box denotes the boundary which gives the highest statistical significance.  On the right, we show the dependence of the significance as a function of the corners of the box.}
\label{fg:scatter-contour-plots}
\end{figure} 
Inclusion of $\met$ associated with the missing neutrino allows even better separation.  In the left plot of Fig.~\ref{fg:scatter-contour-plots}, we show the two-dimensional distribution of $S_{p_T}^\prime$ and $\Delta \phi_{bb,\ell~\met}$.  This distribution shows a high degree of separation between the DPS $Wb\bar b$ and the SPS plus background samples.   To quantify the degree of separation, we define a region in this plane that gives the highest statistical significance.  Its boundary is denoted by the black box in the left panel of Fig.~\ref{fg:scatter-contour-plots}.  Restricting $S_{p_T}^\prime < 0.25$ and $\Delta \phi_{bb,\ell~\met} < 3 \pi/4$, we find a a sample of 154 signal and 103 background events, corresponding to a statistical significance of $S/\sqrt{B} = 15.2$.  

By employing distributions in both $S_{p_T}^\prime$ and $\Delta \phi_{bb,\ell~\met}$, we achieve a better  significance than from $S_{p_T}^\prime$ alone.   By utilizing only $S_{p_T}^\prime$ we obtain a lower significance of 12.7.   In the right plot of Fig.~\ref{fg:scatter-contour-plots}, we show the dependence of the significance on the placement of the box.  As long as the maximum value of $\Delta \phi_{bb,\ell~\met}$ is in the $\pi/2$-$3\pi/4$ range, a statistically significant extraction of DPS $Wb\bar b$ from the other events can be obtained, given our assumed effective cross section $\sigma_{\rm eff} = 12$~mb and luminosity. 

We suggest experimental analyses of $W b \bar{b}$ at the LHC in terms of the two-dimensional distributions presented in this section with the goal to establish whether a discernible DPS signal is  
found.  Assuming success, the $p_T$ dependence of the leading object and other properties of these DPS events can be contrasted with those of the remainder to establish whether the expected properties of DPS are seen.   The enriched DPS event sample can be used for a direct measurement of the effective cross section $\sigma_{\rm eff}$.

\section{Conclusions and Further Work}
\label{sec:conclusions}

In this paper, we investigate the possibility to observe double parton scattering at the early LHC in the 
$p p \rightarrow Wb\bar{b} X \to \ell \nu b\bar{b} X$ process.  Our analysis begins with the basic assumption that $Wb\bar{b}$ production consists of two components: the traditional single parton scattering process and the double parton scattering process where two individual hard scatterings produce the $Wb\bar{b}$ final state, as depicted in the right panel of Fig.~\ref{fg:sps-dps-cartoons}.  

After identifying the most relevant background processes, we pinpoint a set of observables and cuts which would allow for the best separation between the DPS $W b \bar {b}$ signal and the backgrounds (including the SPS $Wb\bar{b}$ process).  To provide the most precise predictions possible, we generate the DPS $W b \bar {b}$ signal event sample,  the SPS $W b \bar {b}$ sample, and the 
dominant  background event samples at next-to-leading order in QCD.  The main obstacles in the extraction of the DPS signal are the backgrounds from 
$t\bar{t}$ production and the SPS $Wb\bar{b}$ component.   The most efficient way to suppress the $t\bar{t}$ background is with an upper cut on the missing transverse energy of the event, since top quark decays result in larger values of $\met$.  

To separate the DPS component of $Wb\bar{b}$ from the SPS component, we find it useful to employ observables which take into account information on the full final state rather than observables which involve one or two particles.  Examples are the $S_{p_{T}}^{\prime}$ variable (defined in Eq.~(\ref{eq:Sptprime})) and the angle ($\Delta \Theta_{b\bar{b},\ell\nu}$) between the two planes defined by the $b\bar{b}$ and $\ell \nu$ systems, respectively.   By displaying the information from these two observables in two-dimensional distributions, we show in Sec.\ref{subsec:2d-plots}
that it is possible to identify distinct regions in phase space where the DPS events reside.  Utilizing cuts on these observables that enhance the DPS $Wb\bar{b}$ sample, we find that the DPS signal can be observed with a statistical significance in the range $S/\sqrt{B} \sim 12 - 15$.

A similar study of the DPS and SPS and background contributions to the $Z b \bar{b}$ final state would be a valuable contribution.  We remark, however, that the NLO calculation of the SPS component of  this final state has not yet been implemented in a numerical code such as POWHEG needed for a differential analysis like ours for $W b \bar{b}$. 

\begin{figure}[h!]
\includegraphics[scale=0.5]{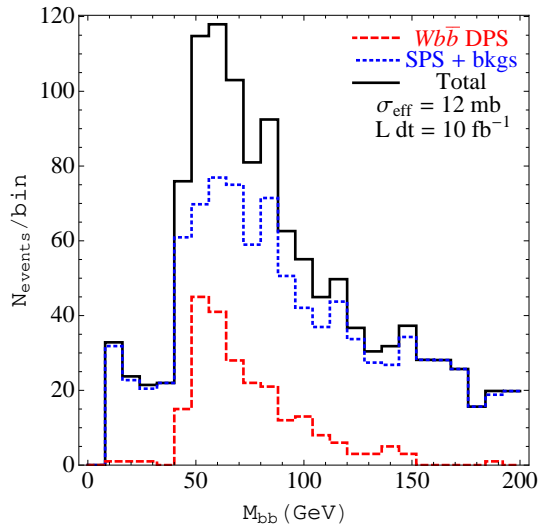}
\caption{The event rate as a function of the invariant mass of the $b\bar{b}$ system using the cuts outlined in the text.}
\label{fg:mbb}
\end{figure}

The focus in the present paper is on establishing double parton scattering as a discernible physics process at LHC energies and measuring the size of its contribution.  Once DPS production of 
$Wb\bar{b}$ is observed, it will be interesting to assess its potential significance as a background in searches for other physics, such as Higgs boson production in association with a $W$ boson (where the Higgs boson decays as $H \to b\bar{b}$), and precise studies of single top quark production where new physics could contribute to the $Wtb$ vertex.   A detailed analysis of either of these channels would require a different set of optimized physics cuts and is beyond the scope of this paper.   We limit ourselves here to showing the $b\bar{b}$ invariant mass distribution for the $\ell\nu b\bar{b}$ final state in Fig.~\ref{fg:mbb}.   These results are for illustration only since they are based on the cuts outlined in this study.  We see that the DPS $W b \bar{b}$ component alters the overall shape of the $b \bar{b}$ mass spectrum, enhancing the small mass region.  This feature is consistent with our earlier observation that the $p_T$ spectrum of leading jets is softer in the DPS component.  In Fig.~\ref{fg:mbb}, we see that  the DPS component contributes primarily in the region below 120 GeV or so.  At face value, it does not seem to pose a hindrance for searches for Higgs bosons in the $HW$ channel.  However, 
$Wb\bar{b}$ DPS could be a significant  background in the search for new particles, with masses in the 50 - 100 GeV range and appearing as resonances in $M_{bb}$, and it should be accounted for in any analysis.

\begin{acknowledgments}

Research in the High Energy Physics Division at Argonne is supported by the U.~S.\ Department of Energy under Contract No.\ DE-AC02-06CH11357.  The research of GS at Northwestern is supported by the U.~S.\ Department of Energy under Contract No.\ DE-FG02-91ER40684.  ELB thanks the Kavli Institute for Theoretical Physics (KITP), Santa Barbara, for hospitality while this research was being completed.  Research at KITP is supported in part by the National Science Foundation under Grant No. NSF PHY05-51164.  
 
\end{acknowledgments}

\end{document}